\documentclass[%
 reprint,
superscriptaddress,
 amsmath,amssymb,
 aps,
]{revtex4-2}

\usepackage{graphicx}
\usepackage{dcolumn}
\usepackage{bm}

\usepackage{hyperref}
\usepackage{mathrsfs}
\usepackage{amsmath}
\usepackage{slashed}
\usepackage{subcaption}
\usepackage{caption}
\usepackage{comment}
\usepackage{multirow}
\usepackage{slashed}
\usepackage[section]{placeins}

\begin{document}

\preprint{APS/123-QED}

\title{Production of Hyperons, Charmed Baryons, and Hadronic Molecule Candidates in Neutrino–Proton Reaction}

	\author{Kai-Sa Qiao}\email{qiaokaisa@itp.ac.cn}
\affiliation{CAS Key Laboratory of Theoretical Physics, Institute of Theoretical Physics, \\
	Chinese Academy of Sciences, Beijing 100190, China}
\affiliation{School of Physics, University of Chinese Academy of Sciences (UCAS), Beijing 100049, China}

\author{Bing-Song Zou} \email{zoubs@mail.tsinghua.edu.cn}
\affiliation{Department of Physics and Center for High Energy Physics, Tsinghua University, Beijing 100084, China}
\affiliation{CAS Key Laboratory of Theoretical Physics, Institute of Theoretical Physics, \\
	Chinese Academy of Sciences, Beijing 100190, China}
\affiliation{Southern Center for Nuclear-Science Theory (SCNT), \\
	Institute of Modern Physics, Chinese Academy of Sciences, Huizhou 516000, China}

\date{\today}

\begin{abstract}
We investigate the production of hyperons, charmed baryons, and potential hadronic molecular states in neutrino–proton ($\bar{\nu}_\mu p$) reaction, a process characterized by a particularly clean final state. Employing effective Lagrangians, chiral perturbation theory, and a hadronic molecular model, we perform theoretical calculations for several relevant channels, including those leading to the formation of the hadronic molecular candidate $(\bar{D}N)$ and $(\bar{D}\Sigma)$. Our results indicate that future neutrino facilities could serve as a complementary platform for exploring exotic baryonic states and provide valuable insights into the dynamics of strong interactions in the strange and charm sectors.

\end{abstract}

\pacs{}
\maketitle


\section{INTRODUCTION}

Over the past two decades, the search for pentaquark states has progressed from early, inconclusive hints to firmly established observations in the heavy-flavor sector~\cite{Guo:2017jvc,Chen:2016qju,Ali:2017jda, Huang:2023jec, Chen:2022asf}. In 2015, the LHCb Collaboration reported two hidden-charm pentaquark-like structures, $P_c(4380)^+$ and $P_c(4450)^+$, observed in the $J/\psi p$ invariant mass spectrum~\cite{PhysRevLett.115.072001}. Subsequent analyses based on higher statistics revealed more refined structures of three narrow resonances $P_c(4312)^+$, $P_c(4440)^+$, and $P_c(4457)^+$\cite{PhysRevLett.122.222001}. These hidden-charm states are located close to the $\Sigma_c \bar D^{(*)}$ thresholds, supporting earlier predictions~\cite{Wu:2010jy,Wu:2010vk,Wang:2011rga,Wu:2012md}  in the picture of hadronic molecules. More recently, evidence for a hidden-charm, strange pentaquark $P_{cs}(4459)$ in the $J/\psi$~$\Lambda$ channel has also been reported~\cite{LHCb:2020jpq}.

On the other hand, Neutrino–proton reactions offer a clean environment for probing hadron spectroscopy~\cite{Wu:2013kla, Ren:2015bsa} and have been extensively investigated in fixed-target experiments over the past decades. With advances in accelerator and detector technologies, modern neutrino experiments can now achieve much higher statistics~\cite{Evans:2013pka, PhysRevD.90.112017, Camilleri2020, MicroBooNE:2025kqo}. For example, the NOMAD experiment at CERN~\cite{ASTIER20023} conducted detailed measurements of inclusive strange-particle production, while the Fermilab MINER$\nu$A experiment~\cite{MINERvA:2004gta} has been designed for precision studies of exclusive strange-channel reactions. The MINER$\nu$A program explicitly aims to perform “precision measurements of exclusive strange-particle production channels near threshold” (e.g.~$\nu_\mu p \to \mu^- K^+ \Lambda$) and to determine hyperon production cross sections and polarizations, thereby enabling searches for pentaquark-like resonances. In summary, current and future neutrino experiments are expected to surpass the old bubble‐chamber datasets by providing high-statistics samples of exclusive $\nu p$ hadronic final states.

In this work, we investigate several representative processes involving hyperons or charmed baryons in the final state, such as $\mu^+K\Lambda$, $\mu^+ \bar D\Lambda$, and $\mu^+\bar DN$, and present their Dalitz plots along with the corresponding total cross sections. We employ effective Lagrangian methods and chiral perturbation theory to estimate their production rates.

Since the tree-level processes discussed above may involve resonant contributions from hadronic molecular states, we also investigate the neutrino-induced production of two molecular candidates, $(\bar{D}N)$~\cite{PhysRevD.105.034028, yamaguchi2022, Yan:2023ttx}and $(\bar{D}\Sigma)$~\cite{Yan:2023ttx,  Yalikun:2021dpk, Wang:2023eng}, denoted as $P_{\bar{c}}$ and $P_{\bar{c}s}$, respectively. In particular, we calculate the cross sections for the two-body processes $\bar{\nu}_\mu p \rightarrow \mu^+ P_{\bar{c}(s)},$ for comparison with the three-body channels.

We aim for the calculations of these representative processes to serve as a preparatory study for exploring their feasibility in future neutrino experiments, thereby providing valuable insights into hadron structure.

This paper is organized as follows. Section~\ref{sec:FORMALISM} presents the theoretical formalism, including the effective Lagrangians and form factors used for the $\nu p$ scattering amplitudes. Section~\ref{sec:NUMERICAL_RESULTS} demonstrates the numerical results for the form factors and computes the cross sections for selected exclusive channels, followed by a discussion of the results. Section~\ref{sec:summary} provides a brief summary and concluding remarks.

\section{FORMALISM}
\label{sec:FORMALISM}
\subsection{Feynman diagrams and Lagrangians}
\begin{figure*}[htbp]
	\centering
	\begin{subfigure}{0.32\textwidth}
		\centering
		\includegraphics[width=\linewidth]{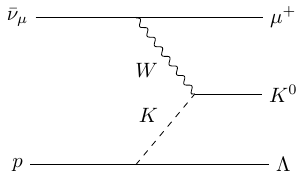}
		\caption{$\bar{\nu}_\mu + p\rightarrow \mu^+ + K^0+ \Lambda$.}
		\label{fig:tree1}
	\end{subfigure}
	\hfill
	\begin{subfigure}{0.32\textwidth}
		\centering
		\includegraphics[width=\linewidth]{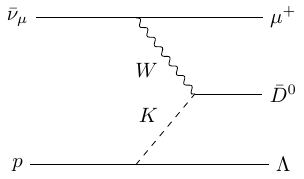}
		\caption{$\bar{\nu}_\mu + p\rightarrow \mu^+ + \bar{D}^0+ \Lambda$.}
		\label{fig:tree2}
	\end{subfigure}
	\hfill
	\begin{subfigure}{0.32\textwidth}
		\centering
		\includegraphics[width=\linewidth]{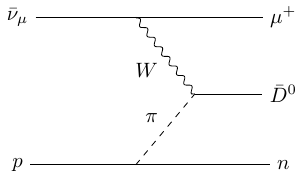}
		\caption{$\bar{\nu}_\mu + p\rightarrow \mu^+ + \bar{D}^0+ n$.}
		\label{fig:tree3}
	\end{subfigure}
	\caption{Feynman diagrams for neutrino–proton scattering leading to three-body final states. Panel (a) shows a hyperon production channel, whereas panels (b) and (c) depict charm production processes.}
	\label{fig:trees}
\end{figure*}

At tree level, we consider three processes for comparison, as shown in Fig.~\ref{fig:trees}. 
We find that the contact terms arising from chiral perturbation theory are numerically much smaller than the meson-exchange contributions, and therefore they are neglected in the following analysis. In Fig.\ref{fig:tree1}, the $W$ boson interacts with $u$ and $d$ quarks, while in Figs.\ref{fig:tree2} and \ref{fig:tree3}, it interacts with $c$ and $s$ quarks. Given that $(\bar{D}N)$ is a candidate for a hadronic molecular state~\cite{yamaguchi2022, PhysRevD.105.034028}, we also consider its production in this process for comparison, as shown in Fig.~\ref{fig:loop1}.

\begin{figure*}[htbp]
	\centering
	\begin{subfigure}{0.32\textwidth}
		\centering
		\includegraphics[width=\linewidth]{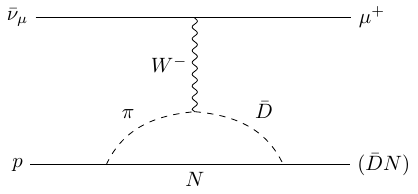}
		\caption{$\bar{\nu}_\mu + p\rightarrow \mu^+  + (\bar{D}N)$.}
		\label{fig:loop1}
	\end{subfigure}
	\hspace{0.1\textwidth}
	\begin{subfigure}{0.32\textwidth}
		\centering
		\includegraphics[width=\linewidth]{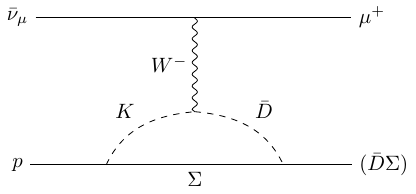}
		\caption{$\bar{\nu}_\mu + p\rightarrow \mu^+ + (\bar{D}\Sigma)$.}
		\label{fig:loop1}
	\end{subfigure}
	
	\caption{($\bar{D}N$) and $(\bar{D}\Sigma)$ molecule production in antineutrino-proton scattering.}
		\label{fig:loop}
	\end{figure*}

For the $(\bar{D}N)$ molecular state, we consider the two isospin configurations $I=0$ and $I=1\,,\,I_3=0$. Their explicit isospin wave functions are
\begin{gather}
	|(\bar{D}N), I = 0\rangle = \frac{1}{\sqrt{2}}\left(|D^- p\rangle - |\bar{D}^0 n\rangle\right)
	\label{eq:isospin0} \\[4pt]
	|(\bar{D}N), I = 1,\, I_3 = 0\rangle = \frac{1}{\sqrt{2}}\left(|D^- p\rangle + |\bar{D}^0 n\rangle\right)
	\label{eq:isospin1}
\end{gather}

For the $(\bar{D}\Sigma)$ molecular state, we restrict ourselves to the $I=1/2$ configuration, which is the most relevant channel for forming a bound or near-threshold molecular state in existing theoretical studies~\cite{PhysRevD.105.034028, Yan:2023ttx}. The corresponding isospin wave function with $I_3=\tfrac{1}{2}$ is given by
\begin{equation}
	|(\bar{D}\Sigma), I=\tfrac{1}{2},\, I_3=\tfrac{1}{2}\rangle
	= \sqrt{\frac{2}{3}}\,|D^- \Sigma^+\rangle
	- \sqrt{\frac{1}{3}}\,|\bar{D}^0 \Sigma^{0}\rangle ,
	\label{eq:isospinDsigma}
\end{equation}

We use effective Lagrangians and the electroweak Lagrangian to describe these processes. At the electroweak interaction vertex, the charged current part of the Lagrangian is given by
\begin{equation}
	\mathscr{L}_c
	= -\frac{g}{\sqrt{2}} \left( \bar{\nu}_l \gamma^\mu\frac{1-\gamma^5}{2}l \right) W_\mu^+ + \text{h.c.}
\end{equation}
Here, $g = e/\sin\theta_w$ is the SU(2) gauge coupling constant, where $\theta_w$ is the weak mixing angle. Its relation to the Fermi constant $G_F$ is given by
\begin{equation}
	\frac{G_F}{\sqrt{2}} = \frac{g^2}{8M_W^2}
\end{equation}
where $M_W$ is the mass of the W boson.

At the hadron level, we use chiral perturbation theory (ChPT)~\cite{Scherer:2012xha} to describe the interactions between hadrons and W bosons. In ChPT, the W boson appears as part of the left-handed current,
\begin{equation}
	l_\mu = -\frac{g}{\sqrt{2}}(W_\mu^+ T_+ + \mathrm{h.c.}),
\end{equation}
where $T_+$ is related to the CKM matrix $V_{ij}$:
\begin{eqnarray}
	T_+ = \begin{pmatrix}
		0 & V_{ud} & V_{us} \\
		0 & 0 & 0 \\
		0 & 0 & 0
	\end{pmatrix}
\end{eqnarray}

The corresponding chiral Lagrangian is
\begin{equation}
	\mathscr{L}_2 = i\frac{F_0^2}{2}\mathrm{Tr}(l_\mu\partial^\mu U^\dagger U) + \cdots
	\label{eq:w-hadron}
\end{equation}
where $F_0$ is the decay constant of the Goldstone bosons in the three-flavor chiral limit. Empirically, the value of $F_\pi$ is $92.4\ \mathrm{MeV}$, and the ratio $F_K/F_\pi \approx 1.2$~\cite{FlavourLatticeAveragingGroupFLAG:2024oxs}. The field U is an $\mathrm{SU}(3)$ matrix defined as $U = \exp(i\phi/F_0)$, where $\phi$ contains the pseudoscalar Goldstone bosons:
\begin{equation}
	\phi =
	\begin{pmatrix}
		\pi^0 + \frac{1}{\sqrt{3}} \eta^0 & \sqrt{2}\pi^+ & \sqrt{2}K^+ \\
		\sqrt{2}\pi^- & -\pi^0 + \frac{1}{\sqrt{3}} \eta^0 & \sqrt{2}K^0 \\
		\sqrt{2}K^- & \sqrt{2}\bar{K}^0 & -\frac{2}{\sqrt{3}} \eta^0
	\end{pmatrix}
\end{equation}

The interaction Lagrangian between baryons and pseudoscalar mesons in ChPT is given by
\begin{equation}
	\mathscr{L}{\phi BB} = -\frac{D}{2F_0} \mathrm{Tr}( \bar{B} \gamma^\mu \gamma_5 \{ \partial_\mu \phi, B \} ) - \frac{F}{2F_0} \mathrm{Tr}( \bar{B} \gamma^\mu \gamma_5 [ \partial_\mu \phi, B ] )
	\label{eq:phiBB}
\end{equation}
where $D = 0.80$ and $F = 0.50$ at tree level~\cite{Borasoy:1998pe}.
The matrix $B$ contains the octet of the $J^P = \frac{1}{2}^+$ baryons and is given by
\begin{equation}
	B=\left(
	\begin{matrix}
		\frac{1}{\sqrt{2}}\Sigma^0+\frac{1}{\sqrt{6}}\Lambda^0 & \Sigma^+ & p\\
		\Sigma^- & -\frac{1}{\sqrt{2}}\Sigma^0+\frac{1}{\sqrt{6}}\Lambda^0 & n \\
		\Xi^- & \Xi^0 & -\frac{2}{\sqrt{6}}\Lambda^0
	\end{matrix}\right)
\end{equation}

One of the tree-level processes (Fig.~\ref{fig:tree3}) involves a hadron pair in the final state, which could potentially form a hadronic molecule. To estimate the likehood of such formation, we also consider the corresponding loop diagram as shown in Fig.~\ref{fig:loop1}. The effective Lagrangian for the hadronic molecule vertex is given by

\begin{equation}
	\begin{aligned}
		\mathscr{L}_{(\bar{D}N)} (x)= &i g_{\bar{D}N}\bar{P}_{\bar{c}}(x) \int d^4y \varPhi(y^2)N(x+w_{\bar{D}N}y)\\
		& \times \bar{D}(x-w_{N\bar{D}}y)+H.c.
	\end{aligned}
	\label{eq.moleculeLagrangian}
\end{equation}
where $P_{\bar{c}}$ denotes the $(\bar{D}N)$ molecular state with quantum numbers $J^P = 1/2^-$. The coupling constants are taken as $g_{\bar{D}N}^{I=0} = 1.68$ and $g_{\bar{D}N}^{I=1} = 2.62$, as determined in our previous work~\cite{PhysRevD.111.056029}. The coupling constant $g_{\bar{D}\Sigma} = 0.66$ is determined using the Weinberg compositeness criterion, corresponding to a binding energy of $0.2~\mathrm{MeV}$~\cite{Yan:2023ttx}. $\omega_{\bar{D}N}$ is a kinematic parameter that reflects the mass ratio between the two constituents of the molecule, defined as
\begin{equation}
	\omega_{ij} = \frac{m_i}{m_i + m_j} ,
\end{equation}
where $m_i$ and $m_j$ are the masses of the constituent particles. $\varPhi(x)$ is a correlation function that characterizes the distribution of the constituent momenta at the vertex. In momentum space, it is defined via a Fourier transform:
\begin{equation}
	\varPhi(x^2)= \int \frac{d^4p}{(2\pi)^4}e^{-ip\cdot x}\widetilde{\varPhi}(-p^2)
\end{equation}
We adopt a Gaussian form with a cutoff parameter $\Lambda$ to model this vertex function:
\begin{equation}
	\widetilde{\varPhi}(p_E^2)  \doteq \exp\left(-\frac{p_E^2}{\Lambda^2}\right) 
\end{equation}
where $p_E$ denotes the Euclidean momentum.

We also introduce two form factors for the exchanged mesons to suppress their off-shell effects during the calculation
\begin{gather}
	f_1(q^2) = \frac{\Lambda_1^4}{\Lambda_1^4 + (q^2 - m_{\mathrm{ex}}^2)^2} \\
	f_2(q^2) = \left( \frac{\Lambda_2^2 - m_{\mathrm{ex}}^2}{\Lambda_2^2 - q^2} \right)^2
\end{gather}
where $q$ is the four-momentum of the exchanged meson, and $\Lambda_1$, $\Lambda_2$ are phenomenological cutoff parameters.

The form factor $f_1$ is used for mesons exchanged in the triangle loop diagram, while $f_2$ is applied in the tree-level diagrams, as shown in Appendix A.

\subsection{Form factor}

Although we have incorporated some form factors, this treatment remains incomplete. The $W$-hadron vertex also requires a form factor to account for the internal structure of mesons. These form factors can be calculated using lattice QCD~\cite{PhysRevD.96.054514,PhysRevD.107.094516,FlavourLatticeAveragingGroupFLAG:2024oxs} or determined from fits to experimental data~\cite{PhysRevD.80.032005,PhysRevLett.121.171803,LINK2005233}. The vector and scalar form factors $f_+(q^2)$ and $f_0(q^2)$ at the $D$-meson vertex can be parameterized as
\begin{equation}
	\begin{aligned}
		\langle P|V_\mu|D \rangle =&\ f_+(q^2)\left(p_{D\mu}+p_{P\mu}-\frac{m_D^2-m_P^2}{q^2}q_\mu\right)\\
		& +f_0(q^2)\frac{m_D^2-m_P^2}{q^2}q_\mu,
	\end{aligned}
\end{equation}
where $P$ represents the final-state mesons $\pi$ or $K$. The form factors in the $q^2$-plane can be expressed using the $z(q^2, t_0)$ expansion, which exhibits rapid convergence:
\begin{equation}
	z(q^2,t_0) = \frac{\sqrt{t_+-q^2}-\sqrt{t_+-t_0}}{\sqrt{t_+-q^2}+\sqrt{t_+ - t_0}}.
\end{equation}
The semileptonic region is given by $m_\ell^2\le q^2 \le t_-$, where $t_- = (M_D-M_P)^2$. A generic form factor contains poles and a branch cut $[t_+,\infty)$ along the real axis, where $t_+ = (M_D+M_P)^2$ is the pair-production threshold. With the choice $t_0 = 0$, the physical region $q^2\in [0,q_{\text{max}}^2]$ maps to $z\in[0,-z_{\text{max}}]$.

A widely adopted parameterization is the BCL form~\cite{BCL2009}:
\begin{align}
	f_0(z) &= \frac{1}{1-q^2(z)/M_{0^+}^2}\sum_{n = 0}^{M-1}b_n z^n,\\
	f_+(z) &= \frac{1}{1-q^2(z)/M_{1^-}^2}\sum_{n=0}^{N-1}a_n\left(z^n-\frac{n}{N}(-1)^{n-N}z^N\right)
\end{align}

The masses $M_{J^P}$ in the denominators represent possible sub-threshold poles. For $D\rightarrow \pi$ transitions, $M_{0^+} = m_{D^{*0}}$ and $M_{1^-} = m_{D^{*}_0}$, while for $D\rightarrow K$ transitions, $M_{0^+} = m_{D^{0}_s}$ and $M_{1^-} = m_{D^{*}_{s}}$. The coefficients $a_n$ and $b_n$ are the series expansion parameters that can be obtained from Ref.~\cite{PhysRevD.107.094516}.

As for the $KKW$ vertex, it is difficult to obtain the form factor from experiments since it does not have a semileptonic decay mode. One approach is to extract it from fits to $\tau^\rightarrow K^-KS\nu\tau$ decay data~\cite{CLEO:1996rit, PhysRevD.98.032010}, and there are phenomenological theoretical analyses for form factor calculations~\cite{Gonzalez-Solis:2019iod}. However, since we are primarily concerned with the order of magnitude of particle production rates, overly detailed form factor structures have minimal impact on the overall scale. Therefore, for simplicity and given our focus on order-of-magnitude estimates, we adopt the VMD (Vector Meson Dominance) model to describe this vertex:
\begin{equation}
	F_V(q^2) = \frac{M_\rho^2}{M_\rho^2 - q^2}
\end{equation} 

In loop calculations, to reduce computational complexity, we evaluate the loop integral contributions in advance and express them as form factors at the $W\textendash p\textendash(\bar{D}N)$ vertex. However, this process can only be completed numerically, and directly using these results in subsequent steps would still be computationally expensive. To address this issue, we employ interpolation functions to handle these numerical results, with details provided in the following section.

The Lagrangian for the $W\text{–}p\text{–}(\bar{D}N)$ vertex can be written as
\begin{equation}
	\begin{aligned}
		\mathcal{L}_{B'BV} =& \bar{B'}_1(g_{B'BV}\gamma_5\gamma^\mu+\frac{f_{B'BV}}{m_1-m_2}\gamma_5\sigma^{\mu\nu}\partial_\nu)V_\mu B_2 \\
		&+ H.c.
	\end{aligned}
\end{equation}
where $B’$ and $B$ denote baryon fields with quantum numbers $J^P = \frac{1}{2}^-$ and $\frac{1}{2}^+$, respectively, and $V$ denotes a vector field, which in this case is the $W$ boson. The couplings $g_{B’BV}$ and $f_{B’BV}$ incorporate the loop-induced form factor, allowing the interaction to be treated effectively as a tree-level vertex.

\section{NUMERICAL RESULTS}
\label{sec:NUMERICAL_RESULTS}
\subsection{Form factor}

In this section, we first introduce the behavior of form factors in the processes under consideration. For the $DKW$ and $D\pi W$ vertices, lattice QCD calculations only consider $q^2>0$ since they are compared with $D$ semileptonic decay data. We extend the range to the $q^2<0$ region, and the results are shown below in Fig.~\ref{fig:semiformfactor}.

\begin{figure}[htbp]
	\centering
	\begin{subfigure}[b]{0.9\linewidth}
		\includegraphics[width=\linewidth]{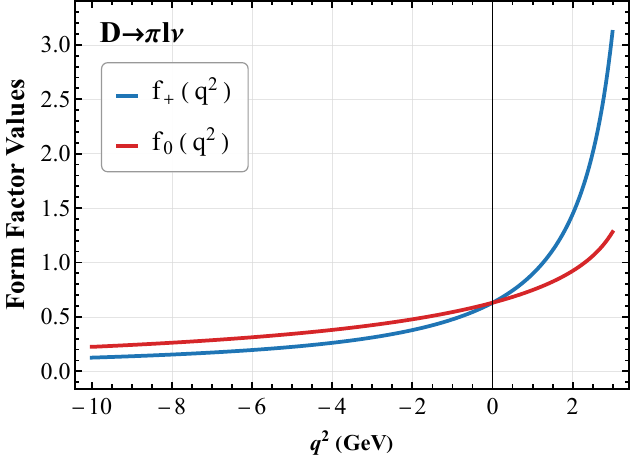}
		\caption{$D\rightarrow \pi \ell \nu$}
		\label{fig:D2Pi}
	\end{subfigure}
	\vspace{0.1cm} 
	\begin{subfigure}[b]{0.9\linewidth}
		\includegraphics[width=\linewidth]{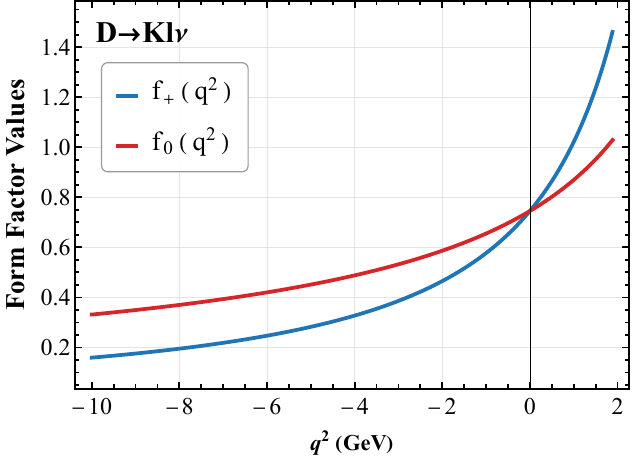}
		\caption{$D\rightarrow K \ell \nu$}
		\label{fig:D2K}
	\end{subfigure}
	
	\caption{Vector and scalar form factors in $D$ meson semileptonic decays: (a)~$D\rightarrow \pi \ell \nu$ and (b)~$D\rightarrow K \ell \nu$.}
	\label{fig:semiformfactor}
\end{figure}

\begin{figure}[htbp]
	\includegraphics[width=1\linewidth]{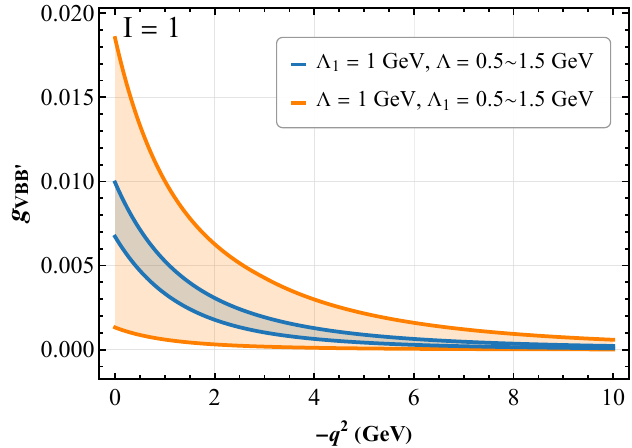}
	\caption{Dependence of the coupling constant $g_{pP_{\bar{c}}W}$ on the cutoff parameters $\Lambda$ and $\Lambda_1$, each varied from 0 to $1~\mathrm{GeV}$. Here, $P_{\bar{c}}$ denotes the $(\bar{D}N)$ state, and the isospin $I=1$ configuration is chosen.}
	\label{fig:formfactocutoff}
\end{figure}

The form factor in the hadronic molecule vertex calculation is relatively complex, as the required integration can only be performed numerically. To simplify its implementation in subsequent calculations—such as those for the tree-level diagrams—we adopt an interpolation-based approximation. Specifically, we first evaluate the integral numerically to obtain a set of data points $\{q^2_i, \Gamma^\mu(q_i^2)\}$. To improve the uniformity of the data distribution, we transform $\Gamma^\mu(q_i^2)$ to $\log[\Gamma^\mu(q_i^2)]$, thereby reducing the relative variation among data points. Next, we construct an interpolation function $f(q^2)$ from the transformed dataset $\{q^2_i, \log[\Gamma^\mu(q^2)]_i\}$. We employed the \texttt{Interpolation} function in \textit{Mathematica}, using the default \texttt{InterpolationOrder}. Finally, in subsequent calculations, we use $\exp[f(q^2)]$ as an efficient representation of the original form factor.
As discussed in the previous section, the form factor in the loop vertex depends on two cutoff parameters,  $\Lambda$ and $\Lambda_1$, whose correlation is illustrated in Fig.\ref{fig:formfactocutoff}. In this plot, one parameter is fixed at $1\mathrm{GeV}$, while the other is varied from $0.5$ to $1~\mathrm{GeV}$.
 
\begin{figure}[htbp]
	\includegraphics[width=1\linewidth]{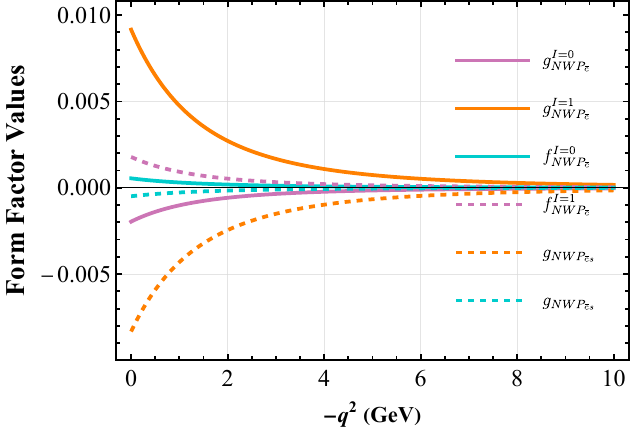}
	\caption{Dependence of the coupling constant $g_{pP_{\bar{c}}W}$ on the cutoff parameters $\Lambda$ and $\Lambda_1$, each varied from 0 to $1~\mathrm{GeV}$. Here, $P_{\bar{c}}$ denotes the $(\bar{D}N)$ state, and the isospin $I=1$ configuration is chosen.}
	\label{fig:couplings}
\end{figure}

From the figure, we can see that the effect of the two cutoffs on the form factor is limited. Therefore, in the subsequent calculations, we set both $\Lambda$ and $\Lambda_1$ to 1 GeV. The $(\bar{D}N)$ molecular states with different isospin values interact differently with their constituents, leading to variations in the corresponding coupling constants. We present these coupling constants to illustrate that they are of the same order of magnitude. As shown in Fig.~\ref{fig:couplings}, the two isospin states have opposite signs, and the coupling constants for $I=0$ are slightly larger than those for $I=1$. 
In contrast, the coupling constant of the $(\bar{D}\Sigma)$ molecular state is about one order of magnitude smaller than those of the $(\bar{D}N)$ states. This suppression mainly originates from its smaller binding energy and the larger mass of the exchanged particle in the loop diagrams.

\subsection{Neutrino–Proton Scattering}
\subsubsection{Dalitz Plot and Invariant Mass Spectrum}

In this section, we first present the Dalitz plot for the process
$ \bar{\nu}_\mu p \rightarrow \mu^+ \bar{D}^0 \Lambda$ . Since the other two processes are similar, we do not show their plots here.

\begin{figure*}[htbp]
	\centering
	\begin{subfigure}{0.32\textwidth}
		\centering
		\includegraphics[width=\linewidth]{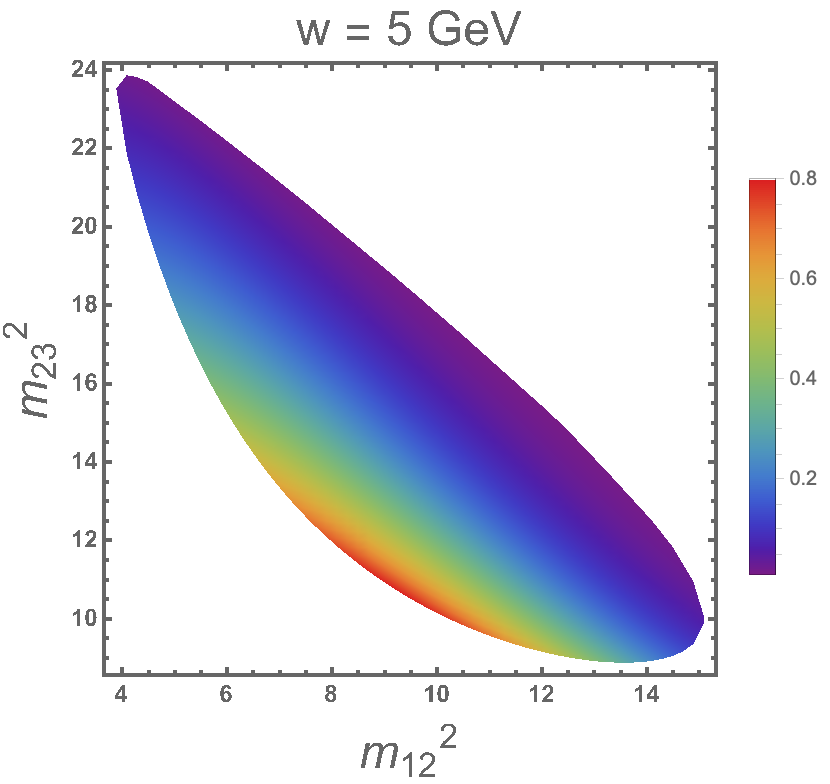}
		\caption{Total energy $w = 5~\mathrm{GeV}$}
		\label{fig:Dalitz5}
	\end{subfigure}
	\hfill
	\begin{subfigure}{0.32\textwidth}
		\centering
		\includegraphics[width=\linewidth]{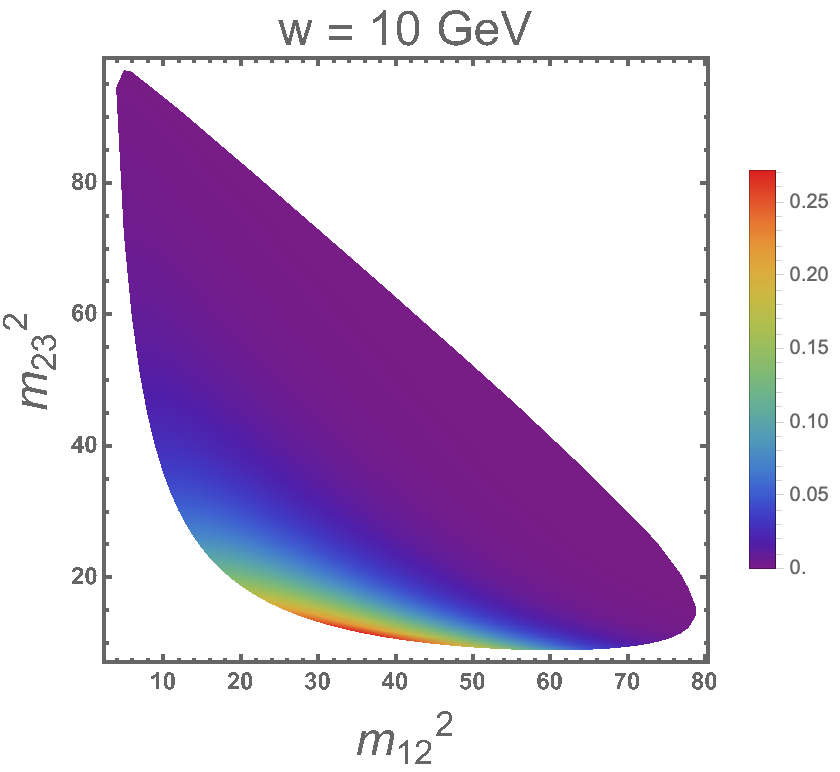}
		\caption{Total energy $w = 10~\mathrm{GeV}$}
		\label{fig:Dalitz10}
	\end{subfigure}
	\hfill
	\begin{subfigure}{0.32\textwidth}
		\centering
		\includegraphics[width=\linewidth]{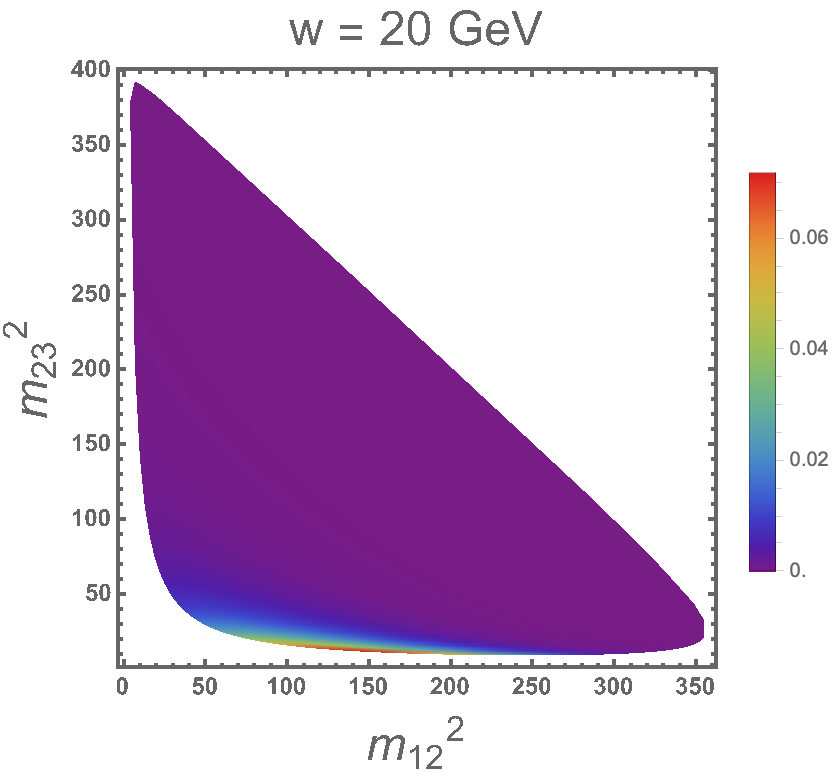}
		\caption{Total energy $w = 20~\mathrm{GeV}$}
		\label{fig:Dalitz20}
	\end{subfigure}
	\caption{Dalitz plots of the process $\bar{\nu}_\mu p \rightarrow \mu^+ \bar{D}^0 \Lambda$ at total energies 
		$w = 5$, $10$, and $20~\mathrm{GeV}$. 
		In this process, $m_1$, $m_2$, and $m_3$ correspond to $\mu^+$, $\bar{D}^0$, and $\Lambda$, respectively. 
		The invariant masses are defined as $m_{ij}^2 = (p_i + p_j)^2$, with the cutoff parameter fixed at 
		$\Lambda_2 = 1~\mathrm{GeV}$. 
		The color scale represents the differential cross section 
		$d\sigma/(dm_{12}^2\, dm_{23}^2)$ in units of 
		$10^{-42}~\mathrm{cm}^2\,\mathrm{GeV}^{-4}$.}
	\label{fig:DalitzPlot}
\end{figure*}

\begin{figure}[htbp]
	\includegraphics[width=0.9\columnwidth]{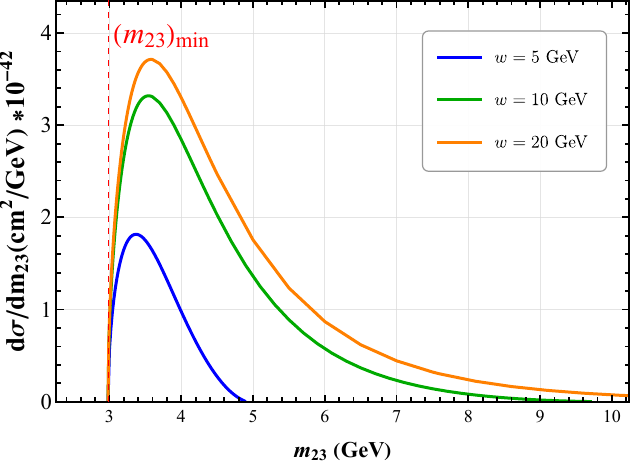}
	\caption{Invariant mass spectra of $\bar D\Lambda$ corresponding to the Dalitz plots in Fig.~\ref{fig:DalitzPlot}, for total energies of 5, 10, and 20 GeV, respectively.}
	\label{fig:IMSDn}
\end{figure}

From Fig.~\ref{fig:DalitzPlot}, we observe that the events are concentrated in the low-$m_{23}^2$ region.
This feature arises from the form factor at the $D K W$ vertex. The horizontal position of the populated region is influenced by the form factor applied to the exchanged $K$ or $\pi$ meson, and therefore depends on the parameter $\Lambda_1$, although the effect is relatively small. As the total energy increases, the dominant event region in the Dalitz plot appears visually narrower; however, the invariant mass spectrum shown in Fig.\ref{fig:IMSDn} reveals that its actual width in $m_{23}$ becomes broader with increasing energy. The broadening rate slows down at higher energies, suggesting that the width will eventually saturate.

In Ref.\cite{Yalikun:2021dpk}, three narrow hadronic molecules of $\bar D\Sigma$ and $\bar D^*\Sigma$ are predicted to exist at energies around 3.05 and 3.19~GeV respectively. They have large decay branching ratio to  $\bar D\Lambda$, hence may be looked for here from the $\bar D\Lambda$ invariant mass spectrum of the reaction $\bar{\nu}_\mu p \rightarrow \mu^+ \bar{D}^0 \Lambda$.

\subsubsection{Cross Section}

In this section, we present the total cross sections corresponding to the processes shown in Figs.~\ref{fig:trees} and \ref{fig:loop1}. As discussed in the previous section, the results depend on the cutoff parameter introduced in the form factor for the exchanged particles. This parameter typically affects the overall magnitude of the final result. Empirically, the cutoff $\Lambda_2$ is usually about $0.4\text{–}1.0\ \mathrm{GeV}$ larger than the mass of the exchanged particle.

\begin{figure}[htbp]
	\includegraphics[width=0.9\linewidth]{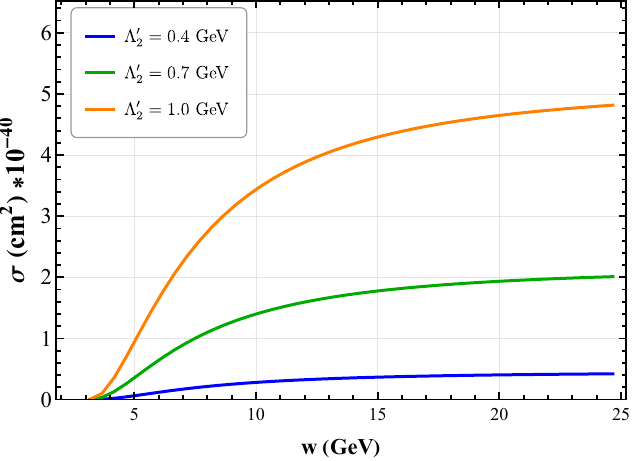}
	\caption{Effect of the cutoff on the process $\bar{\nu}_\mu p \rightarrow \mu^+ \bar{D}^0 \Lambda$. The parameter $\Lambda_2’ = \Lambda_2 - m_\pi$ varies from $0.4$ to $1.0\ \mathrm{GeV}$.}
	\label{fig:cuttofflambda2}
\end{figure}

In Fig.~\ref{fig:cuttofflambda2}, we show three representative curves for $\Lambda_2’$ ranging from $0.4$ to $1.0\ \mathrm{GeV}$ in the process $\bar{\nu}_\mu + p \rightarrow \mu^+ + \bar{D}^0 + \Lambda$, where we define $\Lambda_2’ \equiv \Lambda_2 - m_\pi$ to make the effect more evident. As can be seen, the variation of $\Lambda_2’$ within this range changes the total cross section by roughly one order of magnitude, indicating that the cutoff has only a modest impact on the result.

\begin{figure}[htbp]
	\includegraphics[width=0.9\linewidth]{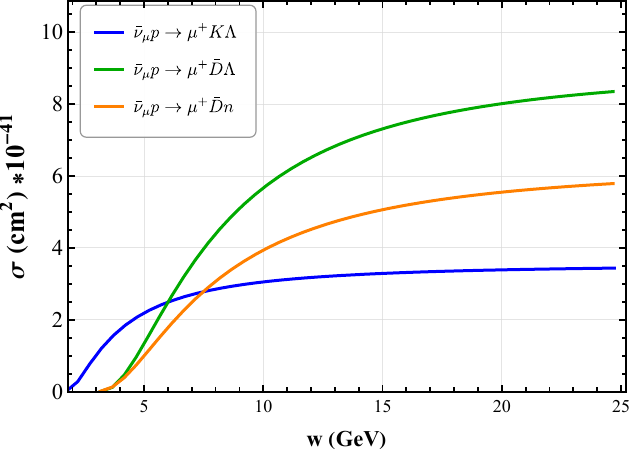}
	\caption{Total cross sections of the three tree-level diagrams from threshold up to a total energy of $w = 25\ \mathrm{GeV}$, with the cutoff parameter fixed at $\Lambda_2 = 1\ \mathrm{GeV}$.}
	\label{fig:treediagrams}
\end{figure}

In Fig.~\ref{fig:treediagrams}, we present the total cross sections for three tree-level processes with the cutoff parameter fixed at $\Lambda_2 = 1\ \mathrm{GeV}$. As shown, all three processes are of the same order of magnitude. The primary reason why $\sigma(\bar{\nu}_\mu p \rightarrow \mu^+ K^0 \Lambda)$ is smaller than $\sigma(\bar{\nu}_\mu p \rightarrow \mu^+ \bar{D}^0 \Lambda)$ is that the form factor at the $KKW$ vertex decreases more rapidly than that at the $DKW$ vertex, even though the former is initially larger at $t = 0\ \mathrm{GeV}^2$. This rapid suppression results in the final cross section being roughly a factor of two smaller.

For the process $\bar{\nu}_\mu p \rightarrow \mu^+ \bar{D}^0 n$, despite the CKM suppression factor $V{cd}/V_{cs} \approx 0.23$, its cross section exceeds that of $\bar{\nu}_\mu p \rightarrow \mu^+ K^0 \Lambda$. This is due to the combined effects of a larger form factor at the $D\pi W$ vertex, larger coupling constants at the $pn\pi$ vertex, and a larger form factor for the exchanged meson when $\Lambda_2 = 1\ \mathrm{GeV}$. These factors collectively account for the behavior observed in the figure.

\begin{figure}[htbp]
	\includegraphics[width=0.95\linewidth]{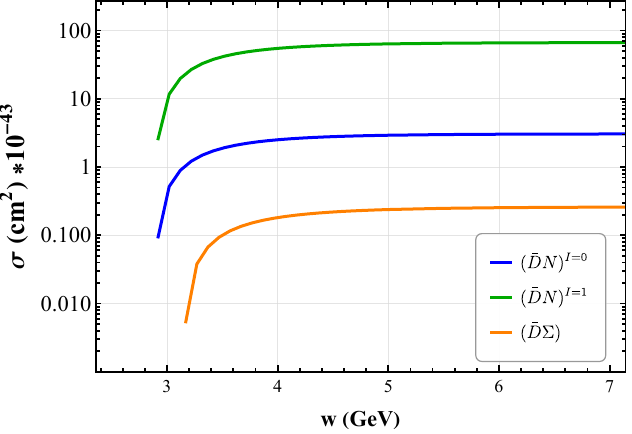}
	\caption{Total cross section of the process $\sigma(\bar{\nu}_\mu p \rightarrow \mu^+ (\bar{D}N))$ and $\sigma(\bar{\nu}_\mu p \rightarrow \mu^+ (\bar{D}\Sigma))$, corresponding to Fig.~\ref{fig:loop}, from threshold up to $w = 7 \ \mathrm{GeV}$.}
	\label{fig:loopdiagram_result}
\end{figure}

Due to the small coupling constants at the hadron vertices in the loop diagrams, the cross sections 
$\sigma(\bar{\nu}_\mu p \rightarrow \mu^+ (\bar{D}N))$ and  
$\sigma(\bar{\nu}_\mu p \rightarrow \mu^+ (\bar{D}\Sigma))$ are significantly smaller than the tree-level results, 
as shown in Fig.~\ref{fig:loopdiagram_result}. 
The results for the two isospin channels, $I = 0$ and $I = 1$, are presented separately. 
In addition, the cross section for the $(\bar{D}\Sigma)$ channel is further suppressed by approximately one order of magnitude compared to that of the $(\bar{D}N)$ channel.

\begin{figure}[htbp]
	\includegraphics[width=0.9\linewidth]{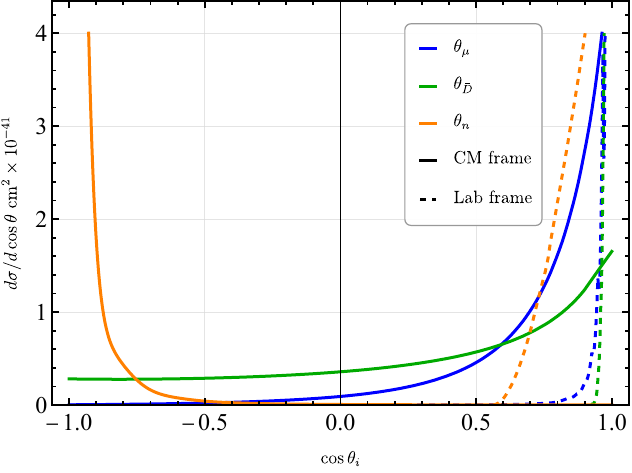}
	\caption{Differential cross sections for $\bar{\nu}_\mu p \rightarrow \mu^+ \bar{D}^0 n$ at $\sqrt{s} = 5 \ \mathrm{GeV}$ in the center-of-mass (solid) and laboratory (dashed) frames. The variable $\theta_i$ denotes the scattering angle of the outgoing particle $i$ ($\mu^+$, $\bar{D}^0$, or $n$) measured with respect to the incident $\bar{\nu}_\mu$ beam direction. }
	\label{fig:differ3}
\end{figure}

\begin{figure}[htbp]
	\includegraphics[width=0.9\linewidth]{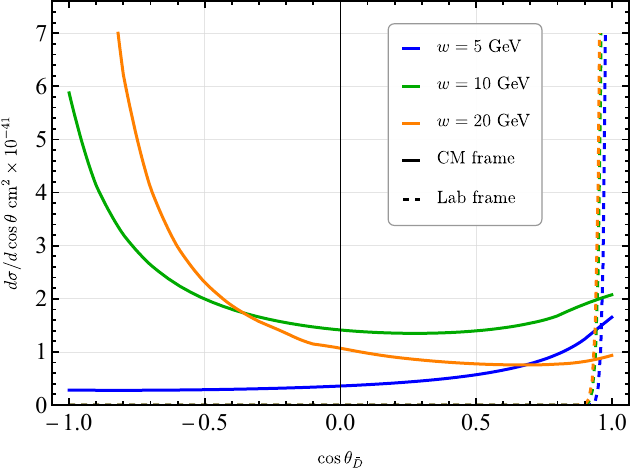}
	\caption{Differential cross sections of the $\bar{\nu}_\mu p \rightarrow \mu^+ \bar{D}^0 n$ process as functions of the $\theta_{\bar{D}^0}$ scattering angle at different center-of-mass energies, illustrating the evolution of the angular distribution. Solid (dashed) lines correspond to the center-of-mass (laboratory) frame.}
	\label{fig:differ1}
\end{figure}

Finally, we present the differential cross section for the process
$\bar{\nu}_\mu p \rightarrow \mu^+ \bar{D}^0 n$. Figure~\ref{fig:differ3} shows the angular differential cross sections of the three final-state particles at a total center-of-mass energy of $\sqrt{s} = 5~\mathrm{GeV}$. 
The solid curves represent the results calculated in the center-of-mass frame, while the dashed curves correspond to those transformed to the laboratory frame. 
As the incident antineutrino is massless, the Lorentz transformation causes the baryon momenta in the laboratory frame to be strongly boosted in the forward direction. 
Consequently, the differential cross sections are concentrated in the region where $\cos\theta \approx 1$. As the energy increases, however, the $\bar{D}^0$ tends to be aligned with the incident proton direction, as illustrated in Fig.~\ref{fig:differ1}. This behavior reflects the kinematic constraints of the reaction and the increasing phase-space availability at higher energies. Such a trend indicates that forward production of heavy mesons becomes more dominant, which is consistent with expectations from hadronic models and may serve as a useful reference for future experimental searches.

\begin{figure}[htbp]
	\includegraphics[width=0.9\linewidth]{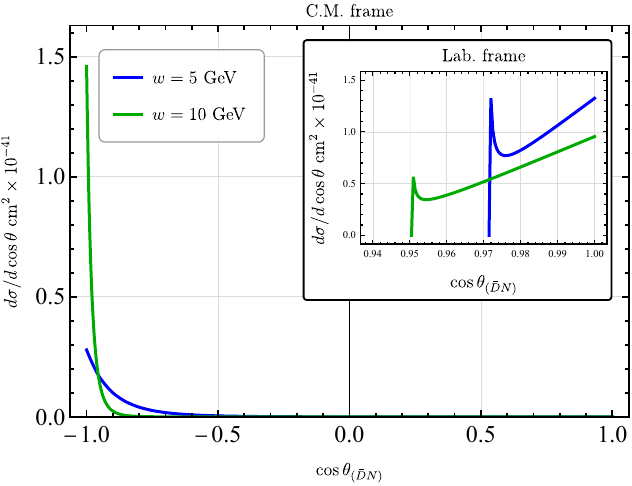}
	\caption{Angular differential cross sections for the process $\bar{\nu}_\mu p \rightarrow \mu^+ (\bar{D}N)$ with isospin $I = 0$. 
Results are shown for total center-of-mass energies $w = 5~\mathrm{GeV}$ and $w = 10~\mathrm{GeV}$. The main panel shows the results in the center-of-mass (C.M.) frame, while the inset illustrates the corresponding distributions in the laboratory (Lab.) frame.
}
	\label{fig:differ_DN_cmlab}
\end{figure}

In addition, we present the angular differential cross sections for the two-body final state process $\bar{\nu}_\mu p \rightarrow \mu^+ (\bar{D}N)$, where $(\bar{D}N)$ denotes a possible molecular state. 
Here, the case with isospin $I = 0$ is shown as a representative example, as shown in Fig.~\ref{fig:differ_DN_cmlab}. It can be seen that, in the center-of-mass frame, the $(\bar{D}N)$ system is predominantly scattered in the backward direction. For the same reason as in the three-body case, due to the massless nature of the incident antineutrino, the Lorentz boost to the laboratory frame results in the $(\bar{D}N)$ momenta being strongly forward-focused, leading to a narrow angular distribution around $\cos\theta \approx 1$.

In particular, the kinematic features observed here provide valuable input for exploring the possible formation of bound states or pentaquark candidates, since the relative momentum distributions of the baryon–meson system play a key role in assessing the likelihood of molecular or multiquark configurations in neutrino–proton scattering.

\section{SUMMARY}
\label{sec:summary}
In this work, we have investigated exclusive hyperon, charmed baryon, and potential hadronic molecular state production in antineutrino–proton scattering within the framework of effective Lagrangians and chiral perturbation theory. Particular attention has been devoted to the $\bar{D}N$ final state, which could serve as a candidate component of hadronic molecular states. By constructing the relevant tree-level and loop amplitudes, incorporating phenomenological form factors, and employing lattice QCD–inspired parameterizations for semileptonic vertices, we have systematically calculated Dalitz plots, invariant mass spectra, and total cross sections for representative channels.

Our results indicate that processes such as $\bar{\nu}_\mu p \to \mu^+ K^0 \Lambda$, $\bar{\nu}_\mu p \to \mu^+ \bar{D}^0 \Lambda$, and $\bar{\nu}_\mu p \to \mu^+ \bar{D}^0 n$ yield cross sections of comparable magnitudes, with differences arising primarily from vertex form factors and CKM suppression effects. The molecular production channels $\bar{\nu}_\mu p \to \mu^+(\bar{D}N)$, $\bar{\nu}_\mu p \to \mu^+(\bar{D}\Sigma)$ are found to be significantly suppressed relative to the tree-level processes, yet still provides a potentially measurable signature at high-statistics neutrino facilities.

These findings highlight the role of neutrino–proton scattering as a complementary probe of hadronic dynamics in both the strange and charm sectors. In particular, future neutrino experiments with enhanced luminosity and detector precision could explore the formation of exotic baryonic states, offering new insights into the structure of hyperons, charmed baryons, and hadronic molecules.

\appendix

\section{Appendixes}

\subsection{Lagrangians and Amplitudes}
Here we provide the explicit Lagrangians and amplitudes we used in our calculation. The hadron Lagrangians from eq.~\eqref{eq:w-hadron},~\eqref{eq:phiBB} are:
\begin{gather}
\mathscr{L}_{pK\Lambda} = \frac{D+3F}{2\sqrt{3}}(\bar{\Lambda}\gamma^\mu\gamma_5p\partial_\mu K^-) + H.c.\\
\mathscr{L}_{WKK} = -\frac{g}{2\sqrt{2}}V_{ud}W_\mu^+(\partial^\mu K^0K^- - K^0\partial^\mu K^-) + H.c. \\
\mathscr{L}_{K^0\pi^\pm W} = -\frac{g}{2\sqrt{2}}V_{us}W_\mu^+(\partial^\mu \bar{K}^0 \pi^- -\bar{K}^0\partial^\mu \pi^-)
\end{gather}
The amplitude in tree diagram Fig.~\ref{fig:tree2} is 
\begin{align}
i\mathcal{M} =&-i(-\frac{g}{\sqrt{2}})[\bar{\nu}(k_1)\gamma^\mu(\frac{1-\gamma_5}{2})\nu(p_1)] \frac{-g_{\mu\nu}+q_{1\mu}q_{1\nu}}{q_1^2-m_W^2} \\
&\times (-\frac{g}{2\sqrt{2}})V_{cx}[f_+(q_1^2)(p_2^\nu+q_2^\nu - \frac{m_{p_2}^2-m_{q_2}^2}{q_2^2}q_1^\nu)\\
& +f_0(q_1^2) \frac{m_{p_2}^2-m_{q_2}^2}{q_2^2}q_1^\nu] \times \frac{1}{q_2^2-m_{q_2}^2}(\frac{\Lambda_2^2-m_{q_2}^2}{\Lambda_2^2-q_2^2}) \\\\
&\times (\frac{D+3F}{2\sqrt{3}})[\bar{u}(p_3)(i\slashed{q}_2)\gamma_5u(k_2)]
\end{align}
for other amplitudes, we can only change their coupling constants from chial peturbation theory. 
\bigskip

\begin{acknowledgments}
	We thank Feng-Kun Guo, Jia-Jun Wu for their useful discussions and valuable comments. This work is supported by the NSFC and the Deutsche Forschungsgemeinschaft (DFG, German Research Foundation) through the funds provided to the Sino-German Collaborative Research Center TRR110 “Symmetries and the Emergence of Structure in QCD” (NSFC Grant No. 12070131001, DFG Project-ID 196253076 - TRR 110), by the NSFC Grant No.11835015, No.12047503, and by the Chinese Academy of Sciences (CAS) under Grant No.XDB34030000.
\end{acknowledgments}

\section*{DATA AVAILABILITY}
No data were created or analyzed in this study.

\bibliography{paperset}

\end{document}